\documentclass[english]{article}
\usepackage[no-natbib-sort]{jheppub}
\usepackage[T1]{fontenc}
\usepackage[latin9]{inputenc}
\usepackage{amsmath}
\usepackage{graphicx}
\usepackage{babel}
\newcommand{\appropto}{\mathrel{\vcenter{
  \offinterlineskip\halign{\hfil$##$\cr
    \propto\cr\noalign{\kern2pt}\sim\cr\noalign{\kern-2pt}}}}}

\newcommand{\overbar}[1]{\mkern 1.5mu\overline{\mkern-1.5mu#1\mkern-1.5mu}\mkern 1.5mu}

\def\be{\begin{equation}}
\def\ee{\end{equation}}
\def\bc{\begin{center}}
\def\ec{\end{center}}
\newtheorem{theorem}{Theorem}
\newtheorem{proposition}[theorem]{Proposition}

\title{Ultrametric identities in glassy models of Natural Evolution}
\author[a,b]{Elena Agliari,}
\author[c,d]{Francesco Alemanno,}
\author[a,b]{Miriam Aquaro,}
\author[b,c]{Adriano Barra}
\affiliation[a]{Dipartimento di Matematica, Sapienza Universit\`a di Roma, Roma, Italy}
\affiliation[b]{GNFM-INdAM, Gruppo Nazionale di Fisica Matematica, Istituto Nazionale di Alta Matematica, Italy}
\affiliation[c]{Dipartimento di Matematica e Fisica, Universit\`a del Salento, Lecce, Italy}
\affiliation[d]{Dipartimento di Matematica, Universit\`a di Bologna, Bologna, Italy}
\emailAdd{adriano.barra@unisalento.it}

\abstract{Spin-glasses constitute a well-grounded framework for evolutionary models.  Of particular interest for (some of) these models is the lack of self-averaging of their order parameters (e.g. the Hamming distance between the genomes of two individuals), even in asymptotic limits, much as like the behavior of the overlap between the configurations of two replica in mean-field spin-glasses. In the latter, this lack of self-averaging is related to peculiar fluctuations of the overlap,  known as Ghirlanda-Guerra identities and Aizenman-Contucci polynomials, that cover a pivotal role in describing the ultrametric structure of the spin-glass landscape. As for evolutionary models, such identities may therefore be related to a taxonomic classification of individuals, yet a full investigation on their validity is missing. In this paper, we study ultrametric identities in simple cases where solely random mutations take place, while selective pressure is absent, namely in {\em flat landscape} models. In particular, we study three paradigmatic models in this setting: the {\em one parent model} (which, by construction, is ultrametric at the level of single individuals),  the {\em homogeneous population model} (which is replica symmetric), and the {\em species formation model}  (where a broken-replica scenario emerges at the level of species). We find analytical and numerical evidence that in the first and in the third model nor the Ghirlanda-Guerra neither the Aizenman-Contucci constraints hold, rather a new class of ultrametric identities is satisfied; in the second model all these constraints hold trivially. 
Very preliminary results on a real biological human genome derived by {\em The 1000 Genome Project Consortium} and on two artificial human genomes (generated by two different types neural networks) seem in better  agreement with these new identities rather than the classic ones.
}


\begin{document}
\maketitle

\section{Introduction} \label{sec:intro}
\subsection{The evolutionary interpretation of spin glasses}
Spin glasses constitute a paradigmatic example of complex system \cite{Parisi1,Parisi2,TonRSB} and their peculiar behavior %in the low-temperature regime 
is often evoked when describing some non-trivial phenomenologies occurring in disparate areas of Science, from several branches of Biology (e.g. neurology \cite{Amit,Neuro2,Neuro3}, genomics \cite{Genomics1,Genomics2}, immunology \cite{Immunology1,Immunology2,BillPnas}, ecology \cite{Ecology1,Ecology2,Sole}) to Sociology \cite{Social1,Social2}, Economics \cite{Economics1,Economics2}, Computer Science \cite{ComputerScience1,ComputerScience2} and more. 

Here we will focus on the framework provided by spin-glass theory to Natural Evolution, which has attracted much interest in the past decades and nowadays represents an insightful and solid branch of modern disordered statistical mechanics.
%
%. As we will see, in this picture genomic randomness plays a crucial role and a natural scaffold is provided by the 
In this picture genomic randomness plays a crucial role; take for instance the {\em adaptive walks} approach, where Natural Evolution is modelled as a two-step stochastic process: $i.$ the genotype of a species undergoes random mutations, $ii.$ its newborns with higher fitness are preserved (see e.g. \cite{Kauffman}). Then, 
as pointed out by Eigen, the compromise between replication efficiency and frequency of mutations in evolutionary dynamics is conceptually close to the compromise between energy minimization and entropy maximization in statistical mechanics, moreover, the {\em error threshold} in the mutation rate in the former mimics the thermal noise of the latter \cite{Franz,Tarazona}. Also, to quantify the genetic variability within a population, we can introduce a proximity measure $q$ between pairs of individuals that plays like an order parameter (mirroring the replica overlap) and, just like in disordered systems, two distinct averages can be implemented, namely the {\em population average} (mirroring the thermal average) and the {\em process average} (mirroring the quenched average) \cite{DP1}. 
Therefore, 
%for models stemming from the adaptive walks approach, 
one can consider the population-average of $q$, which in general depends on time, and inspect whether its average over a long time stretch exhibits vanishing fluctuations. If this is the case we have a {\em self-averaging} structure of the model (in such a way that its main features could be captured by the quasi-species\footnote{Eigen's {\em quasi species} approach (see e.g. \cite{Eigen,Novak}) neglects, by definition, fluctuations and evolution is ruled by deterministic equations reminiscent of reaction kinetics; see also \cite{SciRep1,SciRep2,SciRep3} for a systematic formalization of reaction kinetics via statistical mechanics.} limiting description), and, if not, we have a {\em non-self-averaging} structure, that is a hallmark of complex systems. 

In this context it is also worth recalling Wright's Adaptive Landscape (see e.g. \cite{1932}), Fisher's Fundamental Theorem of Natural Selection and Kimura's Neutral Theory \cite{Kimura} which constitute fundamental steps \cite{Frank}.
Along these lines, the development of a disordered statistical mechanical theory for Natural Evolution was started  by Leuth\"ausser \cite{Lede} and Tarazona \cite{Tarazona} and a spin-glass setting was pursued by Derrida, Higgs, Franz, Peliti, Sellitto, and Serva, just to name a few (see e.g. \cite{DP1,Franz,Derrida2,Derrida3,Peliti2,Peliti-Darwin} and references therein). More specifically, we recognize two classes of models: those where both random mutations and selective pressure are involved (also referred to as \emph{rugged landscape} models) and those where evolution is driven only by random mutations (also referred to as \emph{flat landscape} models). Reference models for the former are the P-spin-glass \cite{Kauffman}, the random energy model (REM) \cite{Franz} and the Hopfield model \cite{Tarazona}, while for the latter we mention the One Parent Model (OPM) \cite{DP1}, the Homogeneous Population Model (HPM) \cite{Peliti2}, and the Specie Formation Model (SFM) \cite{Derrida2,Derrida3}. The OPM is asexuated and its order parameter lacks self-averaging, while its sexuated counterpart, the HPM, is self-averaging, unless a threshold in the similarity between the two genomes that are matching to reproduce is introduced and this case corresponds to the SFM. Notably, in the latter, the presence of a similarity threshold yields a persistent, spontaneous formation and extinction process at the level of species with consequent breakdown of self-averaging.  

The behaviour of the order-parameter fluctuations in spin-glass models has been extensively studied, starting from the fully-connected Sherrington-Kirkpatrick  (SK) model \cite{AC1,Barra0,CG1,GG1}, to its generalizations (see e.g., \cite{Barra2,DeSanctisFranz,FranzLeone,Burioni,PCbook,Dmitry0,ASS1,ASS2,Sollich,Barra3,Bovier1,Bovier2,CG2,CG3,Chatter,More1,More2,Tala} and Sec.~\ref{sec:SK} for more details),
%to its generalization on random and hyper- graphs (known as, respectively, Viana-Bray \cite{Barra2,DeSanctisFranz,FranzLeone} and P-spin models \cite{Burioni,PCbook,Dmitry0}), to its abstract versions such as the Random Overlap Structures (ROSt) by Aizenman, Sims and Starr \cite{ASS1,ASS2} (both fully connected \cite{Barra1} and diluted \cite{Sollich}), to its extensions to neural networks such as the Hopfield model \cite{Barra3}, and to its simpler limits such as the REM \cite{PCbook,Bovier1,Bovier2}, and beyond \cite{CG2,CG3,Chatter,More1,More2,Tala}. In all these cases, which include 
including the rugged landscape models mentioned above. There, the order parameters are proved to be non-self-averaging and their momenta satisfy a class of non-trivial identities known as Ghirlanda-Guerra and Aizenman-Contucci (the latter are actually a family of identities that is a subset of the Ghirlanda-Guerra ones). Thus, in the context of Natural Evolution, the presence of both random mutation and selective pressure seems to be associated to the break-down of self-averaging with the momenta of the order parameter obeying some constraints. The validity of such constraints in the case of flat landscape models is still an open question that deserves attention. 
In fact, we recall that Ghirlanda-Guerra identities played a pivotal role in Panchenko's proof of Parisi ultrametricity in the SK model \cite{Dmitry1,Dmitry2,Dmitry3} and ultrametricity, in turns, covers a key role in Natural Evolution (think for instance at the taxonomic classification in Biology).
In this work we prove analytically for the OPM the validity of a new class of identities and find numerical evidence for their validity also for the SFM for which, instead, classic identities seem to be violated.

\subsection{The harmonic oscillator of spin glasses: Sherrington-Kirkpatrick model} \label{sec:SK}
The SK model \cite{Parisi1,Parisi2,SK0,Dotsenko} is defined in terms of the pairwise Hamiltonian% $\mathcal H_N(\boldsymbol S | \boldsymbol J)$ as
\begin{equation}
\mathcal H_N(\boldsymbol S | \boldsymbol J) = \frac{1}{\sqrt{N}}\sum_{i<j}^{N,N}J_{ij}S_i S_j,
\end{equation}
where the symmetric couplings $\boldsymbol J = \{ J_{ij} \}_{i<j}^{N,N}$  are $\frac{1}{2} N  (N-1)$  i.i.d. random variables sampled from $\mathbb{P}(J_{ij}) = \mathcal{N}(0,1)$\footnote{Beyond sampling from standard Gaussians, the couplings can be drawn with Rademacher entries and the same picture would be the same.} and the interacting units are $N$ Ising spins $\boldsymbol S = \{S_1, ..., S_N \} \in \{ -1, +1\}^N$.
\newline
For  a given inverse temperature $\beta \in \mathbb{R}^+$ and for a quenched coupling setting $\boldsymbol J$, we introduce the Boltzmann-Gibbs measure $\mathcal{P}_{N,\beta}(\boldsymbol S | \boldsymbol J)$, the partition function $\mathcal Z_{N,\beta}( \boldsymbol J)$, and the quenched free-energy $\mathcal F_{N,\beta}$ that read as
\begin{eqnarray}
\mathcal{P}_{N,\beta}(\boldsymbol S | \boldsymbol J) &=& \frac{\exp\left(-\beta \mathcal H_N(\boldsymbol S | \boldsymbol J) \right)}{\mathcal Z_{N,\beta}( \boldsymbol J)},\\
\mathcal Z_{N,\beta}(\boldsymbol J) &=& \sum_{\{ \boldsymbol S \}}^{2^N}\exp\left(-\beta \mathcal H_N(\boldsymbol S | \boldsymbol J) \right),\\
\mathcal F_{N,\beta} &=& \frac{1}{N}\mathbb{E}\ln \mathcal Z_{N,\beta}(\boldsymbol J),
\end{eqnarray}
where the expectation $\mathbb{E}$ is over the possibile realizations of $\boldsymbol J$ drawn from $\mathbb P$. Next, for a generic observable $O(\boldsymbol S)$, we define the following averages 
\begin{eqnarray}
\omega_{N, \beta, \boldsymbol J} (O ) &:=& \sum_{\{ \boldsymbol S \}}^{2^N}O(\boldsymbol S) \mathcal{P}_{N,\beta}(\boldsymbol S | \boldsymbol J)   \\
\langle O \rangle_{N,\beta} &:=& \mathbb{E} [\omega_{N, \beta, \boldsymbol J}  (O)].
\end{eqnarray}
Due to frustration among the spins in the network, once the temperature is lowered beyond a critical one $1/\beta_c$, the free-energy landscape of this system gets spontaneously rugged and minima hierarchically split one into another recursively; consequently, spins tend to freeze in configurations displaying no long-range ferromagnetic-like order. Then, a natural measure of (any) internal organization of the system is a similarity measure between the spin configurations obtained for two replicas of the system characterized by the same realization of disorder $\boldsymbol J$, namely two configurations sampled from the same distribution $\mathcal{P}_{N,\beta}(\boldsymbol S | \boldsymbol J)$. In particular, the (simplest) order parameter is the two-replica overlap 
\begin{equation}
q_{ab}:= \frac{1}{N} \sum_i S_i^{a} S_i^{b}, 
\end{equation}
that is nothing but the normalized scalar product between the configurations, corresponding to two replicas labelled as $a$, $b$, and denoted as $\boldsymbol S^{a}$ and $\boldsymbol S^{b}$. In the high-temperature region, spins behave independently of each other, replica configurations are uncorrelated and the overlap distribution is a Dirac delta peaked at zero, however, beyond $\beta_c$ and in the thermodynamic limit $N \to \infty$, there emerge non-zero values for $q_{ab}$, such that the overlap distribution is a (possibly infinite) sum of Dirac's deltas at these values (i.e. the so-called Parisi plateau), and the whole distribution $\mathcal{P}_{\beta}(q)$ is retained as order parameter.  

%Given a generic function of the spin, we also use the symbols $\omega(f(\sigma))$, $\Omega(f(\sigma))$ and $\langle f(\sigma) \rangle$ to denote the Boltzmann average, the replicated Boltzmann average and the quenched average respectively, defined as  
%$$
%\omega\left(f(\sigma)\right) = \frac{\sum_{\{ \sigma \}}^{2^N} f(\sigma) \exp\left(-\beta H_N(\sigma|J) \right)}{\sum_{\{ \sigma \}}^{2^N}\exp\left(-\beta H_N(\sigma|J) \right)},\ \ \ \  \Omega(f(\sigma))= \omega(f(\sigma)) \times ... \times \omega(f(\sigma)), \ \ \ \ \langle f(\sigma)=\mathbb{E}\Omega\left(f(\sigma)\right) \rangle.
%$$
Thus, as ergodicity breaks down, this model breaks also the permutational invariance among its replicas giving rise to the well-known phenomenon of replica symmetry breaking (RSB): this was suggested  as an ansatz by Giorgio Parisi in the eighties \cite{Parisi1,Parisi2} and then mathematically proved twenty years laters by Francesco Guerra  \cite{Broken} and Michel Talagrand  \cite{Tale} for the expression of free energy and by Dmitry Panchenko \cite{Dmitry1,Dmitry2,Dmitry3} for the hierarchical organization of its valleys, i.e. ultrametricity. Remarkably, Panchenko's proof is significantly based on the peculiar fluctuations of the overlap as summarized by the Ghirlanda-Guerra identities \cite{GG1}, {\em vide infra}. Indeed among the most striking features of the emergent order of the SK model at low temperature lies the spontaneous ultrametric organization of its pure states, resembling taxonomic ordering in Natural Evolution, as for instance captured by the 3-replicas and 4-replicas  overlap joint distributions $\mathcal{P}_{\beta}(q_{12},q_{13})$ and $\mathcal{P}_{\beta}(q_{12},q_{34})$ that read as
\begin{eqnarray}
\mathcal{P}_{\beta}(q_{12},q_{13})=\frac{1}{2}\mathcal{P}_{\beta}(q_{12})\mathcal{P}_{\beta}(q_{13}) + \frac{1}{2}\mathcal{P}_{\beta}(q_{12})\delta \left( q_{12} - q_{13} \right),\\
\mathcal{P}_{\beta}(q_{12},q_{34})=\frac{2}{3}\mathcal{P}_{\beta}(q_{12})\mathcal{P}_{\beta}(q_{34}) + \frac{1}{3}\mathcal{P}_{\beta}(q_{12})\delta \left( q_{12} - q_{34} \right).
\end{eqnarray}
The first expression highlights that, when considering three replicas of the system, it turns out that either two of their overlaps are independent, or they are  identical and these two outcomes happen with the same probability; the second expression confirms that, even when looking at overlaps between two distinct couples of replicas, hence considering four replicas, such a correlation remains strong.
\newline
As a consequence of ultrametricity, along the past two decades a number of constraints on overlap fluctuations in the low temperature regime of spin glasses have been obtained in a mathematically controllable settings  and, among these ensembles of families, the most famous ones are certainly  the Ghirlanda-Guerra identities \cite{GG1},  whose simplest expressions  read as
\begin{eqnarray}
\langle q_{12}^4 \rangle - 2 \langle q_{12}^2 q_{13}^2 \rangle +  \langle q_{12}^2\rangle^2 =0,\\
\langle q_{12}^4 \rangle - 3 \langle q_{12}^2 q_{34}^2 \rangle + 2 \langle q_{12}^2\rangle^2=0,
\end{eqnarray}
as well as their linear counterpart  (where we get rid of $\langle q_{12}^2 \rangle^2$ by substitution in the two equations above),  obtained independently by Aizenman and Contucci \cite{AC1} via stochastic stability (and later with several other techniques \cite{More1,More2,Barra0,CG1}), whence the first identity of the family reads as
\begin{equation}
\langle q_{12}^4 \rangle - 4 \langle q_{12}^2 q_{23}^2 \rangle + 3 \langle 	q_{12}^2 q_{34}^2 \rangle =0.
\end{equation}

Although the SK model remains the archetype of spin glasses, several variations on theme have appeared in the Literature, possibly relaxing its mean-field fully-connected nature. For instance, its version on random graphs (known as Viana-Bray model \cite{VianaBray,Barra2,DeSanctisFranz,FranzLeone}) was studied finding ultrametric fluctuations, that naturally generalize Ghirlanda-Guerra and Aizenman-Contucci identities (and reduce to the latter whenever the coordination number of the graph approached the network size).
and P-spin models \cite{Burioni,PCbook,Dmitry0}
The same holds for models with higher-order interactions (known as P-spin models \cite{Burioni,PCbook,Dmitry0}), even in the diverging number of interactions (known as random energy model, REM \cite{PCbook}), up to extensions as neural networks (e.g., the Hopfield model)  \cite{Barra3}, and beyond \cite{Giorgio,CG2,CG3,Chatter,More1,More2,Tala}.  Further, more abstract representations of the SK model, as for instance the Random Overlap Structures (ROSt) by Aizenman, Sims and Starr \cite{ASS1,ASS2} and its diluted RaMOSt counterpart, also exhibit Ghirlanda-Guerra fluctuations \cite{Barra1,Sollich}.  It is thus rather natural to further inspect the validity of these ultrametric constraints in glassy models of Natural Evolution.

\section{Ultrametric fluctuations in glassy evolutionary models without selective pressure} \label{sec:SG}

Models such as Gardner's P-spin glass \cite{Gardner}, Derrida's REM \cite{REM} or Hopfield's associative memory \cite{Tarazona} have been shown to be plausible models for Natural Evolution under random mutations and selective pressure (see e.g. \cite{Franz,Kauffman,Luca}), also, they are well-known to exhibit overlap fluctuations that respect both the Ghirlanda-Guerra and the Aizenman-Contucci identities.
However, moving to models of Natural Evolution taking place in flat landscapes nothing has been said so far on the validity of these ultrametric constraints, a possible difficulty in answering this question possibly lays in the absence of an Hamiltonian representation for these models. In the following we inspect the three best-known models in this context, that is the One Parent Model (OPM, that is a model for asexual reproduction exhibiting, by construction, RSB on the scale of single progenies), the Homogeneous Population Model (HPM, that is a basic model for sexual reproduction, where reproduction may involve two parents regardless their genetic distance and it is replica-symmetric) and the Species Formation Model (SFM, that is much as the previous model where, crucially, a threshold in string similarity for dating is introduced and the latter turns the evolution of the model to be RSB and at the level of species rather than single genomes).  
\newline
We will show that for the OPM both the Ghirlanda-Guerra and the Aizenman-Contucci identities are violated and we prove the existence of another family of identities that is instead respected. Extending the same analysis on the HPM returns a rather simple scenario where all the identities are trivially respected (as anticipated since the model is replica-symmetric). Next, we tested (numerically) all the three families of ultrametric constraints on the SFM: a finite-size-scaling analysis suggests that they are expected to hold in the suitable limits (i.e., the infinite genome limit and large population limit), with the new class of identities being the ones minimally violated by the finite size effects. Driven by this last finding, we close this section by inspecting whether these constraints are fulfilled on actual genomes, focusing on a sample of the biological human genome and two artificial genomes and we find that the scenario depicted for the SFM is preserved also in these realistic settings.

The simplifying assumptions that we preserve along the paper are those of the original manuscripts (see e.g., \cite{DP1}), namely 
\begin{itemize}
\item
while Evolution takes place the population size is preserved and set equal to $M$; 
\item
each individual $a \in \{1,..., M \}$ is represented by a string of $N$ bits $\{ S_1^{a},\ S_2^{a},\ ..., S_N^{a} \}$, with $N$ constant during the evolutionary process, which can be interpreted as the genome of the individual $a$\footnote{Actually, using a generic $N$-bits vectors allows us to map the string from a binary alphabet to the natural one for the problem under study (e.g., a quaternary one when dealing with the four DNA bases adenine (A), cytosine (C), guanine (G), and thymine (T)) such that, in general, the sequences $\{ S_i^{a}(t) \}_{i=1}^N$ can represent  bases of a nucleic acid sequence, amino acids in a protein, alleles in a genome, etc.};
\item
the genome are subjected to mutations and we will focus on point-mutations\footnote{While real world mutation can include insertion and deletions \cite{BillPnas}  and more complex randomness, the theoretical advantage of single mutations  is that a Markov process in the genome space driven by these mutation has symmetric transitions rates as if -say- genotype A is one-step away from genotype B, then also genotype B is one-step away from genotype A. Further, by the empirical counterpart there is a confirm \cite{Dati1} that the bulk of mutations in human genomes is point-like.} that happen at constant mutation rate (among different generations) and independently of a given locus (i.e., the unit of the genome that mutates): we thus associate one-to-one to each genotype  a phenotype\footnote{In models with selective pressure the latter is used to evaluate the fitness of that genotype such that the higher its fitness the larger the number of its offsprings, but this does not happen in flat landscapes}. 
\item
the dynamics is parallel: at each iteration all the individuals in the populations are removed and replaced by their offsprings.
\end{itemize}
With these simplifications the state of the population at given time $t$ can be described by specifying the genome of all the individuals $\{ S_i^{a}(t) \}_{i=1}^N$. The natural measure of genetic distance between two individuals $a$ and $b$ is the Hamming distance 
\begin{equation}
d_{ab}=\frac{1}{2}\sum_{i=1}^N|S_i^{a} - S_i^{b}| = \frac{N}{2}(1- q_{ab}), 
\end{equation}
where $q_{ab}$ is the overlap between the genomes of the individuals $a$ and $b$ and mirrors the overlap between the spin configurations of two replicas. Analogously, the $M \times M$ matrix $q$ evaluated at a given time $t$ provides a snapshot of the population structure at that time.
%\begin{equation}
%q_{\alpha,\beta}=\frac{1}{N}\sum_{i=1}^N S_i^{\alpha} S_i^{\beta} = 1 - 2d_{\alpha,\beta}/N.
%\end{equation}
Interestingly, it can be proved that, in the $N \to \infty$ limit, the three flat landscape models under consideration can be simulated by directly looking at the evolution of $q$ rather than dealing with the set of genomic sequences \cite{Derrida2,Derrida3}.

\subsection{The One Parent Model (OPM)} \label{sec:OPM}

In the OPM studied by Derrida and Peliti \cite{DP1} we consider a population $\Omega$, made up of a fixed number, $M$,
of individuals reproducing synchronously and asexually, whose genome at generation $t$ is encoded 
by a $N$-bits vector $\boldsymbol{S}^{a}(t) \in\{-1,+1\}^{N}$ for $a=1,...,M$.
At each generation $t$, all the individuals are removed, and a new
generation is formed by offsprings of the previous individuals. 
%We randomly extract the $M$ individuals in $\Omega$ at time $t$ which
%will form the new generation. 
More precisely, each individual $a \in \Omega$ is randomly associated to a parent $G(a) \in \Omega$ and the genome $\boldsymbol{S}^{a}(t)$
is taken identical to that of its parent $\boldsymbol{S}^{G(a)}(t-1)$ at
the previous generation $t-1$ except for random mutations, as specified by
%
%\begin{equation}\label{mutation}
%S_{i}^{a}(t)=\begin{cases}
%S_{i}^{G_{t}(a)}(t-1) & \text{with probability }\frac{1}{2}(1+e^{-2\mu})\\
%-S_{i}^{G_{t}(a)}(t-1) & \text{with probability }\frac{1}{2}(1-e^{-2\mu}).
%\end{cases}
%\end{equation}
\begin{equation}\label{mutation}
\mathbb{P}_1[S_i^a(t) = \pm S_i^{G(a)} (t-1)]= \frac{1}{2}(1 \pm e^{-2\mu}),
\end{equation}
where $\mu \in \mathbb R^+$ tunes the mutation probability; the subscript highlights ``1'' that we are comparing individuals separated by one generation and, in the following, the expectation related to $\mathbb{P}_1$ shall be referred to as $\mathbb E_1$.
%and the overlap $q^{\alpha\beta}$ between two individuals
%$\alpha$ and $\beta$ : 
%\begin{equation}
%q^{\alpha\beta}=\frac{1}{N}\sum_{i}S_{i}^{\alpha}S_{i}^{\beta}
%\end{equation}
%is preferred w.r.t the Hamming distance (as standard in glassy statistical mechanics) as a convenient measure of the genoma variability.
As for the mapping $a \to G(a)$, it is assumed that $G(a)$ is chosen independently and uniformly in $\Omega$ for each individual and at each generation. Therefore, for any individual $a \in \Omega$, its ancestors over the previous $t$ generations are given by the sequence $\{G(a), G^2(a), ..., G^t(a)\} = \Gamma_t(a)$. 
%\begin{equation}
%G_{t}(a)\sim \mathcal U \left(\{1,\cdots,M\}\right)
%\end{equation}
%Moreover, the expression can be properly generalized to $\mathbb P_{t}$ when comparing $S_{i}^{G_{t}(a)}(0)$ and $S_{i}^{a}(t)$.

As remarked in Sec.~\ref{sec:intro}, analogously to spin glasses, we have two averages: at each generation $t$ we can take the average of any quantity involving
the individual of the whole population $\Omega$ ($\textit{population average}$
$\langle\cdot\rangle$) but, as this quantity may fluctuate even for an
infinitely large population $\Omega$ according to the particular mapping sequence $(\Gamma_{t})$ which has taken place, we should consider also the average of these quantities taken over all possible
realizations of the reproduction process $(\textit{process average}$
$\overline{~\cdot~}$). 
Crucially, the process average can be obtained by averaging
over the temporal unfolding of the process for a sufficiently long
time stretch, since the time sequence ($\Gamma_{t}$) of mappings belonging
to different time intervals are independent, but this has to be done with care, properly inspecting the typical decorrelation time of the stochastic process (an analysis that we perform in the next subsection).

\begin{figure}
\begin{centering}
\includegraphics[scale=0.65]{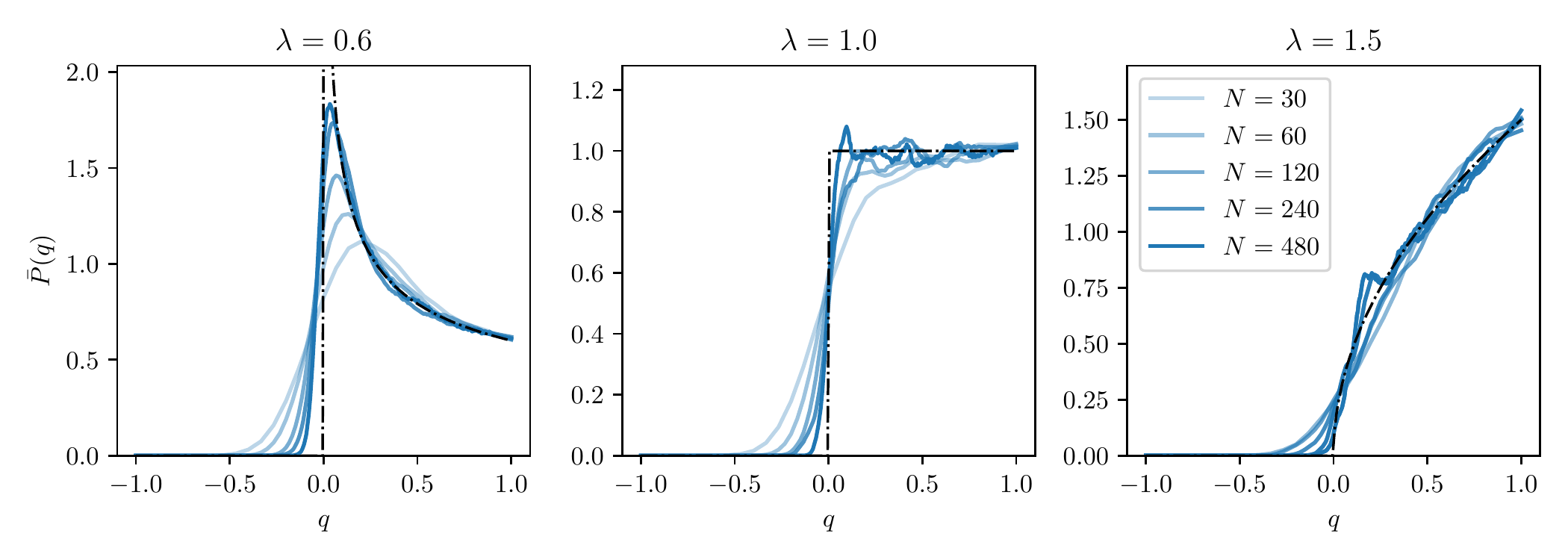}
\par\end{centering}
\caption{\label{FiguraUno} Numerical estimate for the overlap probability density function $\bar{P}(q)$ for a population of $M=50$ individuals averaged over
$10^5$ generations. Several genome sizes are tested and depicted in different colors as explained in the legend, while
the dashed curve corresponds to the theoretical probability distribution given by \eqref{p_q}. 
As expected, if $\lambda < 1$ mutations are likely and the overlap concentrates on values $q \approx 0$, if $\lambda > 1$ mutations are rare and the overlap  concentrates on values $q \approx 1$, while if $\lambda = 1$ the probability distribution becomes uniform. }
\end{figure}

Specifically, at generation $t$, the population average, that we denote with $\langle q \rangle_t$, can be obtained by means of the following 
\begin{equation}
\langle q \rangle_t  = \int q  P(q,t)dq,
\end{equation}
where
\begin{equation}
P(q,t) = \frac{1}{\binom{M}{2}} \sum_{a<b} \delta (q_{ab}(t) - q). 
\end{equation}
Thus, $\langle q \rangle_t$ fluctuates in time about a mean value that we denote with $\overline {\langle q \rangle }$ and which can be expressed as 
\begin{equation}
\overline{\langle q\rangle}=\int q\bar P(q)dq, 
\end{equation}
where $\bar P$ is the overlap distribution averaged over time. 
It can be proven \cite{DP1} that, in the limit $N\gg1$, the time-averaged overlap distribution
depends only on the parameter $\lambda:=\frac{1}{4M\mu}$ and it is
\begin{equation} \label{p_q}
\bar P(q)=\begin{cases}
\lambda q^{\lambda-1} & 0<q\leq1\\
0 & \text{otherwise}
\end{cases}
\end{equation}
such that
\begin{equation} \label{eq:mediona}
\overline{\langle q\rangle}=\int q\bar P(q)dq=\frac{\lambda}{\lambda+1}.
\end{equation}
Notice that for $\lambda<1$ the distribution is peaked at $q=0$, for $\lambda=1$
the distribution is uniform in the interval $0<q\leq1$, and as $\lambda$
exceeds $1$ the peaks is at $q=1$.
As shown in Figure \ref{FiguraUno}, the agreement between theoretical predictions and simulation outcomes is pretty good already for relatively small sizes and the overlap is better and better as $N$ is made larger.  Remarkably, the broad distribution of the overlap $q$ highlights the non-self-averaging nature of the order parameter in the OPM. Indeed, even in the infinite genome-size limit, one has \cite{DP1}
\begin{equation}
\overline{\langle q \rangle^2}-\overline{\langle q\rangle}^2 \neq 0.
\end{equation}

\subsubsection{Exponential decay of correlations for efficient sampling}

If we let the system evolve for a time $t\gg1$, the last generation will be made up of individuals with a unique common ancestor with probability one in the asymptotic limit \cite{DP1}, hence it is possible to find an expression for the decay of genome-correlations between  individuals at a given time and their common ancestor, as a function of time. Let us start evaluating the expectation value of the overlap between a parent $S^{G(a)}(t)$ and the corresponding offspring at $t+1$:
\begin{eqnarray}
q_{\Delta t=1}&=&\mathbb{E}_1 [ \frac{\boldsymbol S^{G(a)}(t)\cdot \boldsymbol S^{a}(t+1)}{N} ] =\frac{1}{N}\sum_{i=1}^{N} \mathbb{E}_1[S_{i}^{G(a)}(t) S_{i}^{a}(t+1) ]\\
&=&\frac{1}{N}\sum_{i=1}^{N} \{ 1 - 2 \mathbb P_1 [S_{i}^{a}(t+1)  = - S_{i}^{G(a)}(t)] \}\\ &=& 1 - 2 \mathbb P_1 [S_{i}^{a}(t+1) = - S_{i}^{G(a)}(t)] = e^{-2\mu},
\end{eqnarray}
where in the last line we exploited the fact that loci are independent.
We can demonstrate that the expectation value of the overlap between a
parent and the corresponding offspring at $t+\Delta t$
is
\begin{equation}
q_{\Delta t}=\mathbb{E}_{\Delta t}  [\frac{S^{G^{\Delta t}(a)}(t)\cdot S^{a}(t+\Delta t)}{N} ] =e^{-2\mu\Delta t},
\end{equation}
where the average $\mathbb E_{\Delta t}$ is performed over the distribution $\mathbb P_{\Delta t}$ which generalizes \eqref{mutation}.
We prove this by induction: first we observe that $q_{\Delta t=1}=e^{-2\mu}$
is true, next we assume that $q_{\Delta t}=e^{-2\mu\Delta t}$
is true for any $\Delta t$ and we check that this is sufficient to ensure that also $q_{\Delta t+1}=e^{-2\mu(\Delta t+1)}$ is true.
In the following, in order to lighten the notation, we shall we set $t=0$ without loss of generality and we shall drop the superscript labelling the individual without any loss of information: the individual we are referring to is $a$ or its ancestor  at the generation specified by time dependence of $\boldsymbol S$.

\begin{figure}
\centering{}\includegraphics[scale=0.7]{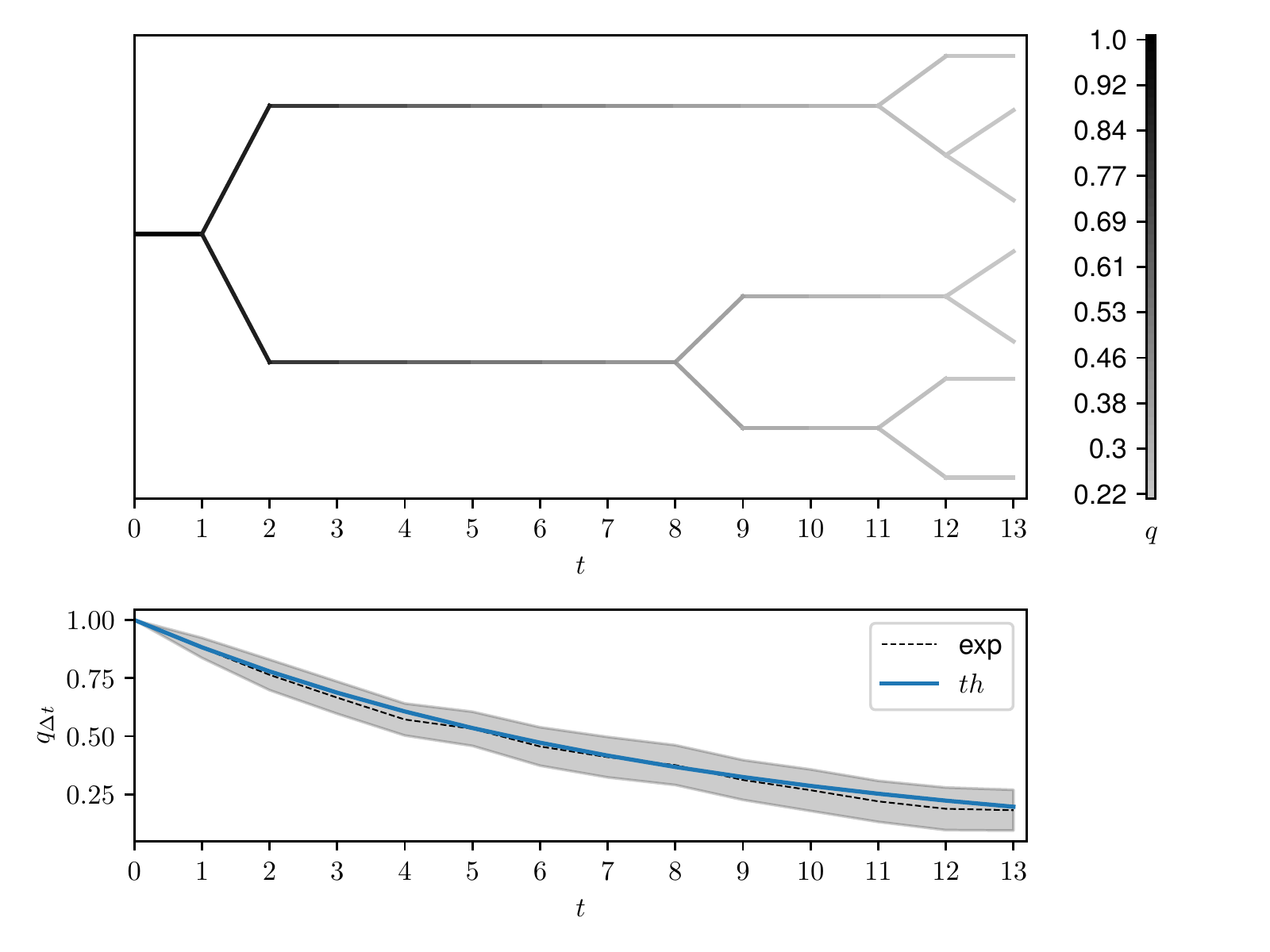}\caption{\label{tree}Upper panel: number of individuals generated by the ancestor as a function of the generation; in this particular evolution, all the individuals present at time $t=13$ turn out to stem from the same common ancestor and other branches that have not survived are omitted in this plot. The colormap highlights the overlap between the various individuals and the ancestor. Lower panel: time decay of the overlap between the ancestor and its offsprings; the numerical estimate (solid line) is consistent with the theoretical estimate $e^{-2\mu t}$ (dashed line); the shadow represents three standard deviations evaluated by repeating the process 100 times.}
\end{figure}

Let use observe that $q_{\Delta t}$ can be written as 
\begin{equation}
q_{\Delta t}=  \frac{1}{N}\sum_{i=1} \{ 1 - 2 \mathbb P_{\Delta t} [ S_{i}(\Delta t)  =-  S_{i}(0) ] \} \label{qdt}
\end{equation}
and, analogously, $q_{\Delta t+1}$ can be written as
\begin{equation}
q_{\Delta t+1}=\frac{1}{N}\sum_{i=1}\left \{ 1-2 \mathbb P_{\Delta t +1}[S_{i}(\Delta t+1)=-S_{i}(0)] \right \}.\label{overight}
\end{equation}
By the law of total probability we can write
\begin{eqnarray}
\nonumber
\mathbb P_{\Delta t +1}[S_{i}(\Delta t+1)&=&-S_{i}(0)]=\mathbb P_{\Delta t}[S_{i}(\Delta t)=S_{i}(0)] ~ \mathbb P_{1 }[S_{i}(\Delta t+1)=-S_{i}(\Delta t)]\\
 &+&\mathbb P_{\Delta t}[S_{i}(\Delta t)=-S_{i}(0)] \mathbb P_1[S_{i}(\Delta t+1)=S_{i}(\Delta t)].
\end{eqnarray}
Recalling that
\begin{equation}
\mathbb P_1[S_{i}(\Delta t+1)= - S_{i}(\Delta t)]=\frac{1-e^{-2\mu}}{2}
\end{equation}
we reach
\begin{equation}
\mathbb P_{\Delta t +1 }[S_{i}(\Delta t+1)=-S_{i}(0)]=\frac{1-e^{-2\mu}}{2}+e^{-2\mu} \mathbb P_{\Delta t}[S_{i}(\Delta t)=-S_{i}(0)].
\end{equation}
By direct substitution of the last equation into (\ref{overight})
we get 
\begin{equation}
q_{\Delta t+1}=\frac{e^{-2\mu}}{N}\sum_{i=1} \left \{1-2 \mathbb P_{\Delta t}[S_{i}(\Delta t)=-S_{i}(0)] \right \}.
\end{equation}
and, recalling the definition of $q_{\Delta t}$ given in (\ref{qdt}), the last equation gets
\begin{equation}
q_{\Delta t+1}=\frac{e^{-2\mu}}{N}\sum_{i=1} \left[1-2 \mathbb P_{\Delta t }(S_{i}(\Delta t)=-S_{i}(0))\right]=e^{-2\mu}q_{\Delta t}=e^{-2\mu(\Delta t+1)}.
\end{equation}
In figure \ref{tree}, the  exponential decay of the overlap between the ancestor and its offsprings as a function of time is shown along with the related family tree.

\subsubsection{A new class of ultrametric identities}
As stressed in Sec.~\ref{sec:OPM}, in the OPM the genome overlap is non self-averaging and $\overline{\langle q^2 \rangle} -  \overline{\langle q \rangle}^2 \neq 0$. However, following spin-glass theory, one may wonder whether other kinds of relation hold among overlap momenta.
As recalled in Sec.~\ref{sec:SG}, in the SK and many variations of its, the Ghirlanda-Guerra and Aizenman-Contucci identities are preserved even under replica-symmetry-breaking. Here, to inspect the validity of these identities we study numerically  the following quantities
\begin{equation}
\varepsilon_{\textrm{GG}}=\overline{\langle q_{12}^{4}\rangle}- 2 \overline{\langle(q_{12}q_{13})^{2}\rangle} + \overbar{\langle q_{12}^{2}\rangle^{2}}\end{equation}
\begin{equation}
\varepsilon_{\textrm{AC}}=\overbar{\langle q_{12}^{4}\rangle}-4\overbar{\langle(q_{12}q_{13})^{2}\rangle}+3\overbar{\langle q_{12}^{2}q_{34}^{2}\rangle}
\end{equation}
\begin{equation}
\varepsilon_{\textrm{SA}}=\overline{\langle q_{12}^{4}\rangle}-\overline{\langle q_{12}^{2}\rangle^{2}},
\end{equation}
where $\varepsilon_{\textrm{GG}, \textrm{AC}, \textrm{SA}}$ are interpreted as a measure of possible violation. 
Remarkably, since our inspection is only based on numerics, at finite population size and along a finite time span, in order to verify if non-null values of $\varepsilon_{\textrm{GG}, \textrm{AC}, \textrm{SA}}$ are intrinsic or, rather, stem from finite-size effects, we will perform a finite-size-scaling: if the extent of $\varepsilon$ is non-decreasing by increasing the system size, we will have a signature for the breakdown of the related identity.

%Following spin glass theory, we expect that  in the thermodynamic limit, both  are preserved, while self-averaging breaks down, i.e. to the lowest order 
%\begin{eqnarray}
%&& \lim_{N \to \infty} ( \langle q_{\alpha,\beta}^2 \rangle -  \langle q_{\alpha,\beta}\rangle^2) \neq 0, \\ 
%&& \lim_{N \to \infty} ( \langle q_{\alpha,\beta}^4 \rangle -  \langle q_{\alpha,\beta}^2\rangle^2) \neq 0, \\ 
%\end{eqnarray}
%Clearly at finite volumes, of both genome length and individuals, we expect all these constraints to be mildly violated and a finite size scaling argument should show if by increasing these volumes such violations diminish or grow.
%\newline

\begin{figure}[tb]
\centering{}\includegraphics[scale=0.7]{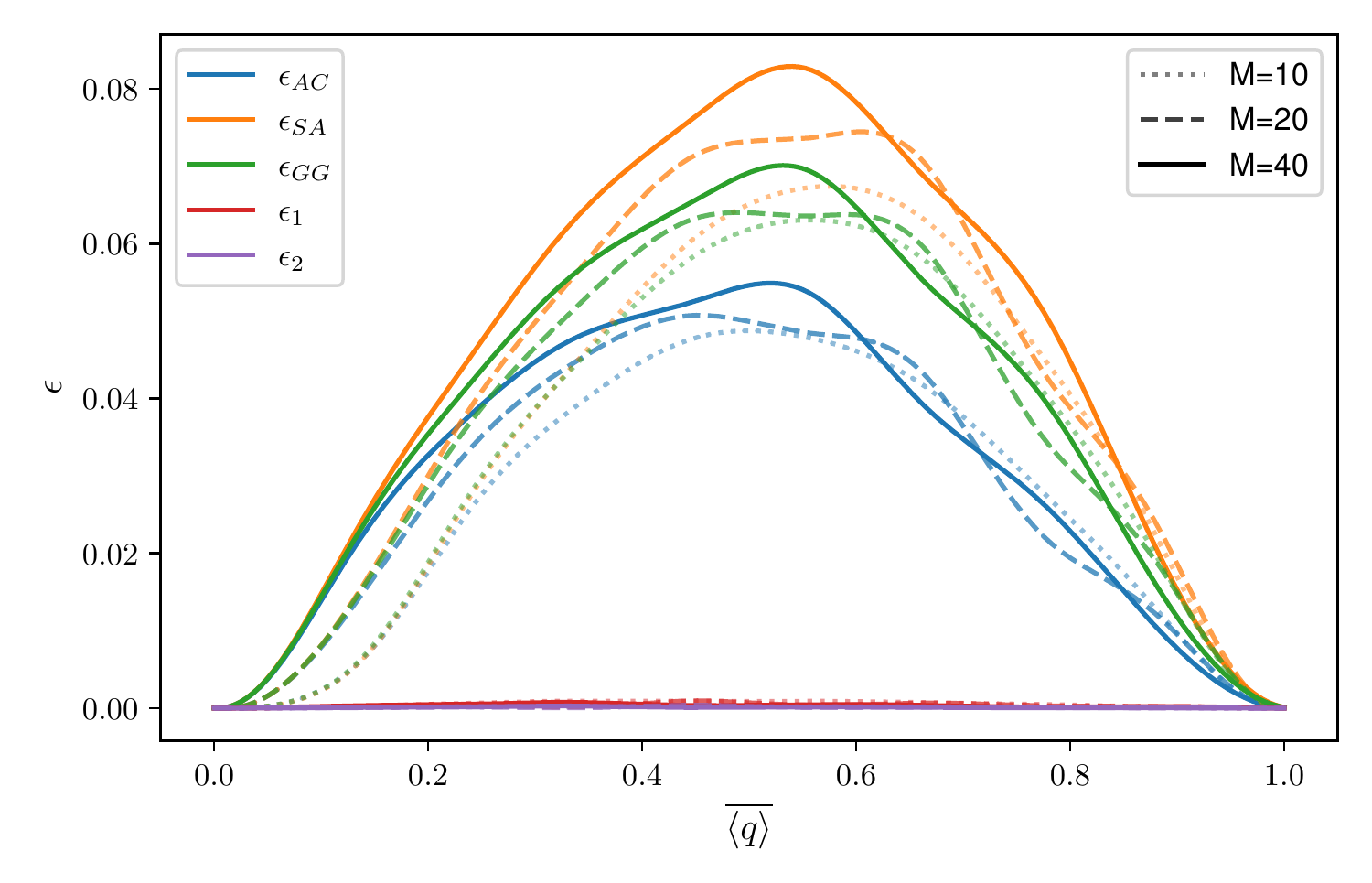}\caption{\label{OPMcurves} Errors measured for the various ultrametric identities as a function of the mean overlap for the OPM and for different values of $M$; the process average is taken over a run of $10^2 \times M$ generations. Note that, as $M$ (and the time span, accordingly) grows, the error on self-averaging ($\varepsilon_{\textrm{SA}}$), on Ghirlanda-Guerra identities ($\varepsilon_{\textrm{GG}}$)  and on Aizenman-Contucci polynomials  ($\varepsilon_{\textrm{AC}}$) does not decrease, while the errors $\varepsilon_{1},\ \varepsilon_2$ are robustly vanishing. }
\end{figure}

Beyond these quantities, we can inspect the time-averaged joint probability density $\bar P(q_{12},q_{23},q_{13})$ of the overlaps
$q$, in the infinite genome limit $N\to\infty$, that is known \cite{DP1} and reads as,
\begin{equation}
\begin{cases}
\bar P(q_{12},q_{23},q_{13})=\frac{\lambda^{2}}{2}\theta(q_{23}-q_{12})\delta(q_{12}-q_{13})q_{12}^{\lambda-1}q_{23}^{2\lambda-1}+\text{Perm}(1,2,3) & q_{12},q_{13},q_{23}\in(0,1]\\
0 & \text{otherwise}
\end{cases}
\end{equation}
looking for possible relations among overlap momenta.
In particular, we find that in the thermodynamic limit $N\to\infty$ 
\begin{equation}
\overline{\left\langle q_{12}^{K}\right\rangle }=\frac{\overline{\left\langle q_{12}\right\rangle }}{K+\overline{\left\langle q_{12}\right\rangle }(1-K)}=\frac{\lambda}{\lambda+K},~\forall K\in\mathbb{N}, \label{q_moments}
\end{equation}
which generalizes \eqref{eq:mediona} and, by a direct calculation, we also find
\begin{equation}\label{qq}
\overline{\left\langle q_{12}^{\alpha}q_{13}^{\alpha}\right\rangle }=\frac{\lambda^2 \left(5 \alpha^2+8 \alpha \lambda+3 \lambda^2\right)}{(\alpha+\lambda)^2 (2\alpha+\lambda) (2\alpha+3 \lambda)}, ~\forall \alpha \in \mathbb R^+.
\end{equation}
Now, by merging \eqref{q_moments} and \eqref{qq}, we obtain the following relation, that plays as a new generator of overlap constraints for this model
\begin{equation}\label{general}
\overline{\langle q_{12}^{2\alpha}\rangle}+\beta\overline{\langle q_{12}^{\alpha}q_{13}^{\alpha}\rangle}-(1+\beta)\frac{\overline{\langle q_{12}^{\alpha}\rangle}\,\overline{\langle q_{12}^{2\alpha}\rangle}\,\overline{\langle q_{12}^{2\alpha/3}\rangle}}{\overline{\langle q_{12}^{\frac{\alpha}{6}\left(5+\sqrt{\frac{25\beta+1}{\beta+1}}\right)}\rangle}\,\overline{\langle q_{12}^{\frac{\alpha}{6}\left(5-\sqrt{\frac{25\beta+1}{\beta+1}}\right)}\rangle}}=0.
\end{equation}
Indeed, the above equation constitutes an infinite family of relations which hold for the OPM. % and can be seen as the equivalent of the Ghirlanda-Guerra and Aizennman-Contucci relations for spin glasses.
%\newline 
In particular, due to the non-integrability of the overlap momenta with negative power, we must have $\beta\geq -1/25, \alpha\geq 0$  which fulfils  the condition $5-\sqrt{\frac{25\beta+1}{\beta+1}}\geq0$ ensuring that the moments of the overlap are well defined. 
\newline
As an example, if we set $\alpha=2$ and $\beta=1/7$ in equation \eqref{general} we get
\begin{equation}\label{ep1}
\overline{\langle q_{12}^{4}\rangle}+\frac{1}{7}\overline{\langle q_{12}^{2}q_{13}^{2}\rangle}-\frac{8}{7}\frac{\overline{\langle q_{12}^{2}\rangle}\,\overline{\langle q_{12}^{4}\rangle}\overline{\langle q_{12}^{4/3}\rangle}}{\overline{\langle q_{12}^{7/3}\rangle}\overline{\langle q_{12}\rangle}}=:\varepsilon_{1},
\end{equation}
then, if we set $\alpha=2$ and $\beta=1/2$ in equation \eqref{general}, we get
\begin{equation}\label{ep2}
\overline{\langle q_{12}^{4}\rangle}+\frac{1}{2}\overline{\langle q_{12}^{2}q_{13}^{2}\rangle}-\frac{3}{2}\frac{\overline{\langle q_{12}^{2}\rangle}\,\overline{\langle q_{12}^{4}\rangle}\overline{\langle q_{12}^{4/3}\rangle}}{\overline{\langle q_{12}^{8/3}\rangle}\overline{\langle q_{12}^{\frac{2}{3}}\rangle}}=:\varepsilon_{2}
\end{equation}
where we introduced $\varepsilon_1$ and $\varepsilon_2$ to measure possible failures of these relations. 
As shown in figure \ref{OPMcurves}, both $\varepsilon_{1}$ and $\varepsilon_{1}$ are numerically found to vanish for the OPM. However, we stress that equations \eqref{ep1} and \eqref{ep2} are just two examples of equalities since there is an infinite family of relationships   which are satisfied by the OPM and that can be obtained by varying $\alpha \in [0,+\infty)$ and $\beta\in[-1/25,+\infty)$ in equation \eqref{general}.
In Figure \ref{OPMcurves} we also show numerical evidence that self-averaging is broken (as expected by construction) and that nor the Ghirlanda-Guerra identities neither the Aizenman-Contucci polynomials seem to hold. 
%Indeed a finite size scaling argument for these quantities does not show any sign of convergence toward zero for their relative errors $\epsilon$.

%%%%
%%%%

\subsection{The Homogeneous Population Model (HPM)}
Serva and Peliti \cite{Peliti2} investigated a natural extension of the OPM, namely a two-parents model where parents can mate regardless their genome proximity; as a result of this feature, the long-time limit population is homogeneous (whence the name given to model) and, consequently, the model exhibits replica-symmetry in such a way that all the ultrametric constraints become trivial identities. 
\newline
\begin{figure}
\centering{}\includegraphics[scale=0.7]{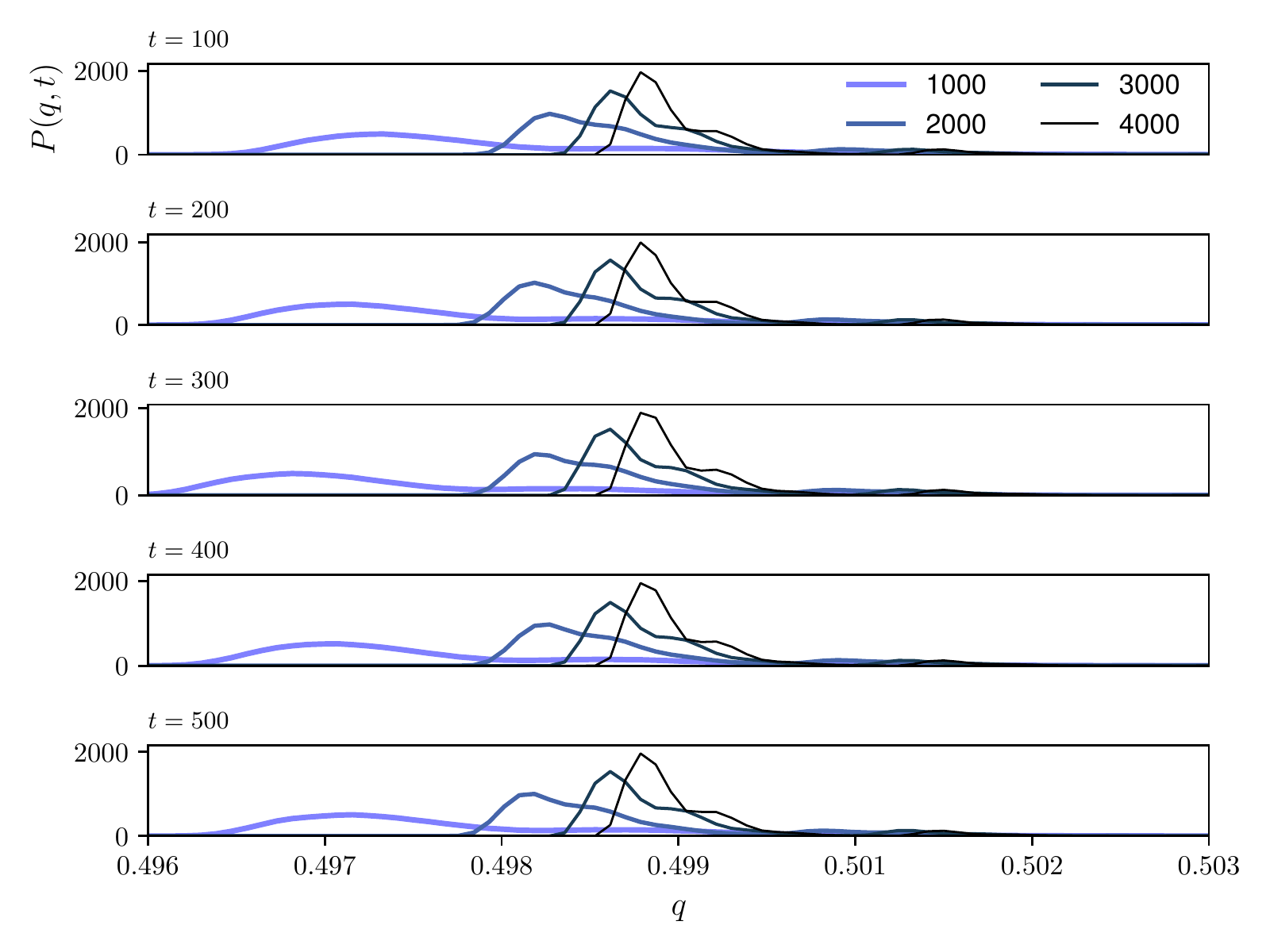}\caption{\label{all} Distribution of the overlap $P(q,t)$ of the HPM at $\lambda=1.0$ and for different values of $M$ as shown in the legend. As $M$ is made larger and larger $P(q,t)$ gets monomodal and peaked at $\langle q \rangle_t=1/2$, thus highlighting a replica symmetric behavior of the specie evolution in the HPM.}
\end{figure}
In the HPM, at each generation $t$, each individual $a$ has two distinct parents $G_{1}(a)$ and $G_2(a)$ chosen at random from the previous generation. Each spin $S_{i}^{a}$ is inherited from either $G_{1}(a)$ or $G_2(a)$ with equal probability, with the same probability of faithful copy or mutation as in equation \eqref{mutation}.
In the OPM model if the overlap between the parents $G(a)$ and $G(b)$ of two individuals, $a$ and $b$, is $q_{G(a)G(b)}$ then the expectation value of the overlap of $a$ and $b$ is 
\begin{equation}
q_{ab}=e^{-4\mu}q_{G(a)G(b)}.
\end{equation}
If $N$ is infinite this becomes a deterministic rule for updating the overlap matrix. There is an equivalent rule for updating the overlap matrix for the HPM in the limit $N\to\infty$. The pair of spins $S_{i}^{a}S_{i}^{b}$ is inherited from one of the four combinations of parents of the two individuals with equal probability, therefore
\begin{equation}
q_{ab}=\frac{e^{-4\mu}}{4}(q_{G_1(a)G_1(b)}+q_{G_2(a)G_1(b)}+q_{G_1(a)G_2(b)}+q_{G_2(a)G_2(b)})
\end{equation}
with $q_{aa}=1$ always. It can be proven that the variance of $\langle q \rangle_t$ vanishes in the limit $M\to\infty$, thus $\langle q \rangle_t$ is self-averaging in the HPM, in particular $\lim_{M\to\infty}\overbar{\langle q \rangle} = \frac{\lambda}{1+\lambda}$. 
\begin{figure}
\centering{}\includegraphics[scale=0.7]{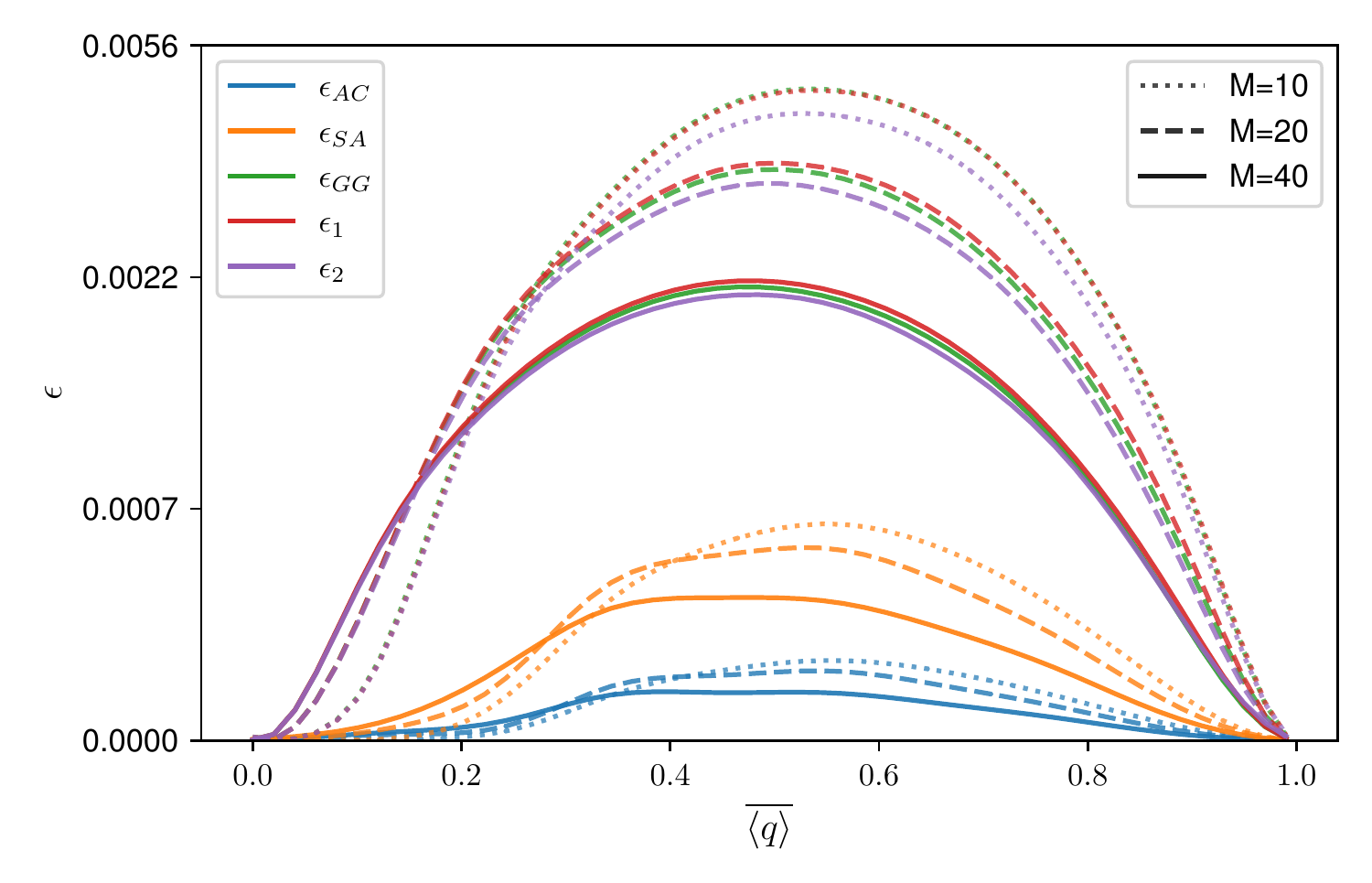}\caption{\label{HSMcurves} Errors measured for the various ultrametric identities as a function of the mean overlap for the HPM and for different values of $M$; the process average is taken over a run of $10^2 \times M$ generations Note that, as $M$ grows, all the errors depicted  tend to approach the horizontal axes as expected since the overlap is self-averaging in this model.}
\end{figure}
Figure \ref{all} shows results about the overlap distribution for long simulations of the HPM model with $\lambda=1$ and, accordingly, $\lim_{M\to\infty} \overline {\langle q \rangle} =1/2$:  the various rows show the overlap distribution $P(q,t)$ sampled at different times and by inspecting its variance as a function of $M$ it can be shown that it scales as $1/M$ hence it is expected to disappear for large population size $M$ such that $P(q,t)$ gets concentrated around $q=1/2$, showing that self-averaging of the overlap is respected hence proving that the model behaves in a replica symmetric manner. Consistently, in figure \ref{HSMcurves} we show that the errors on the ultrametric identities approaches zero as $M$ is made larger and larger. 
 \begin{figure}
\centering{}\includegraphics[scale=0.75]{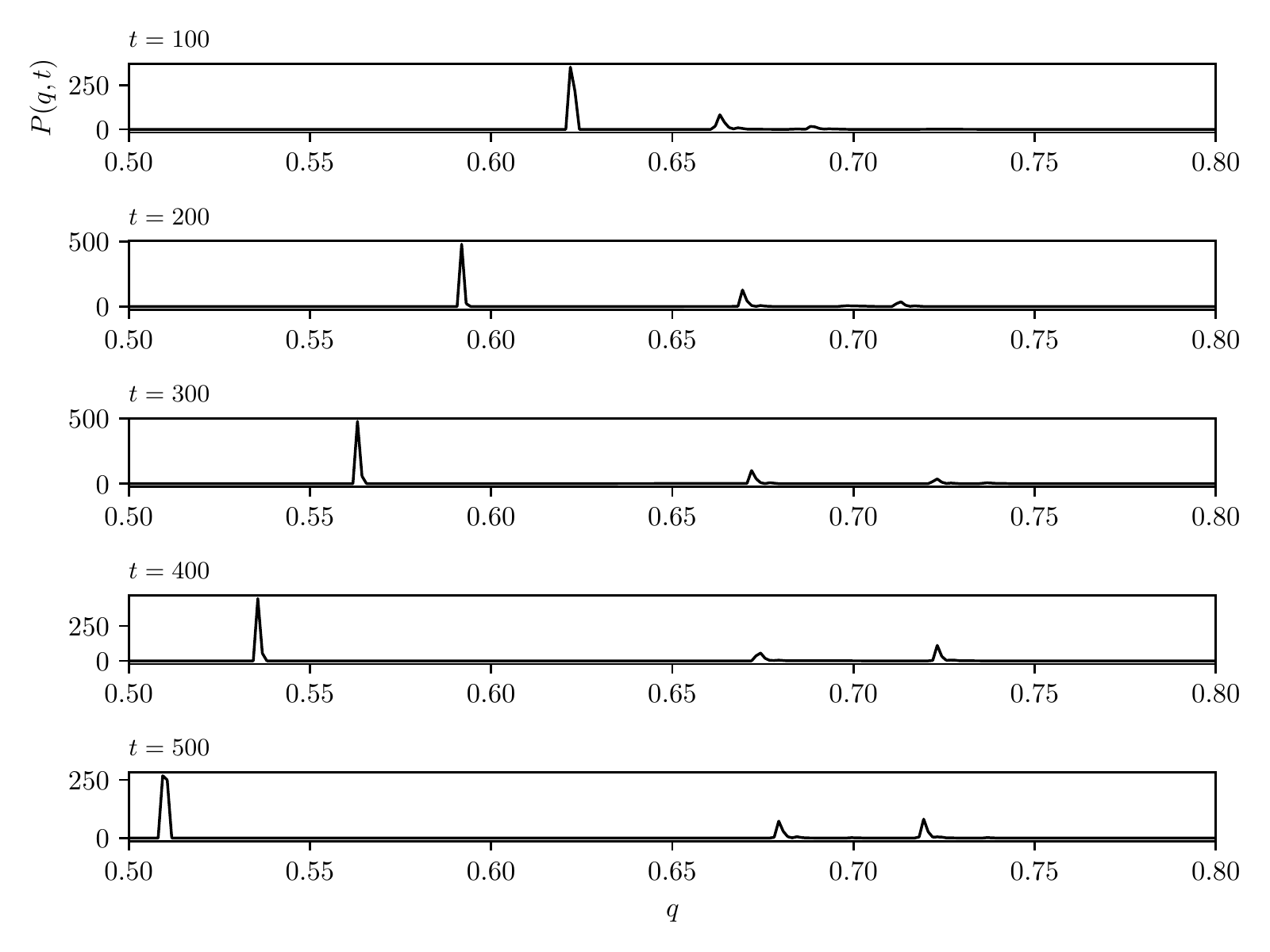}\caption{\label{SFM} Distribution of the overlap $P(q,t)$ of the SFM for $\lambda=1.0,q_{min}=0.65,M=2000$. Note that, at difference w.r.t. the previous replica-symmetric HPM, here there actually are new specie formation and extinction as the presence of several peaks in the $P(q,t)$ evidences. Further, beyond the big one on the left (that is exponentially collapsing toward zero as time goes on), the erratic presence of these small peaks confirms that the  SFM gives rise to an evolutive scenario strongly resembling the paining by replica-symmetry-breaking of the low temperature spin glasses.}
\end{figure}

\subsection{The Specie Formation Model (SFM)}

The SFM introduced by Higgs and Derrida \cite{Derrida2,Derrida3} is  nothing but the two-parents model of Serva and Peliti with  a threshold for mating $q_{min}$. Specifically, it is defined in the same ways as the HPM earlier, except that the first parent $G_1(a)$ of individual $a$ is chosen randomly from the previous generation whereas the second $G_2(a)$ is chosen only from those individuals having an overlap $q_{G_1(a)G_2(a)}$ greater than a cutoff value $q_{min}$\footnote{If no such a second parent is available then a new first parent is randomly selected.}. In the absence of a cutoff, thus in the case of HPM, there is a natural mean value of the overlap $\overbar{\langle q \rangle }= \lambda/(1+\lambda)$, in such a way that, if the SFM $q_{min}>\lambda/(1+\lambda)$, the system is highly perturbed by the introduction of the threshold and it can never reach its natural equilibrium state:  $\pi\acute{\alpha}\nu\tau\alpha\;\grave \rho \varepsilon \widetilde{\iota}$  as in the low temperature regime of spin glasses.
\newline
A corroboration of this picture appears in figure \ref{SFM} where we show the distribution of the overlap $P(q,t)$ at different times:  the peaks appearing above the threshold $q_{min}>0.65$ correspond to the overlaps of the new species that have formed (and their disappearance to their extinction)  while the large peak below the threshold exponentially collapses toward zero (since such values of the overlap are lower than $q_{min}$ no interbreeding is possibile between them and, consequently, the peak must be vanishing).
%%%%
\begin{figure}[tb]
\centering{}\includegraphics[scale=0.7]{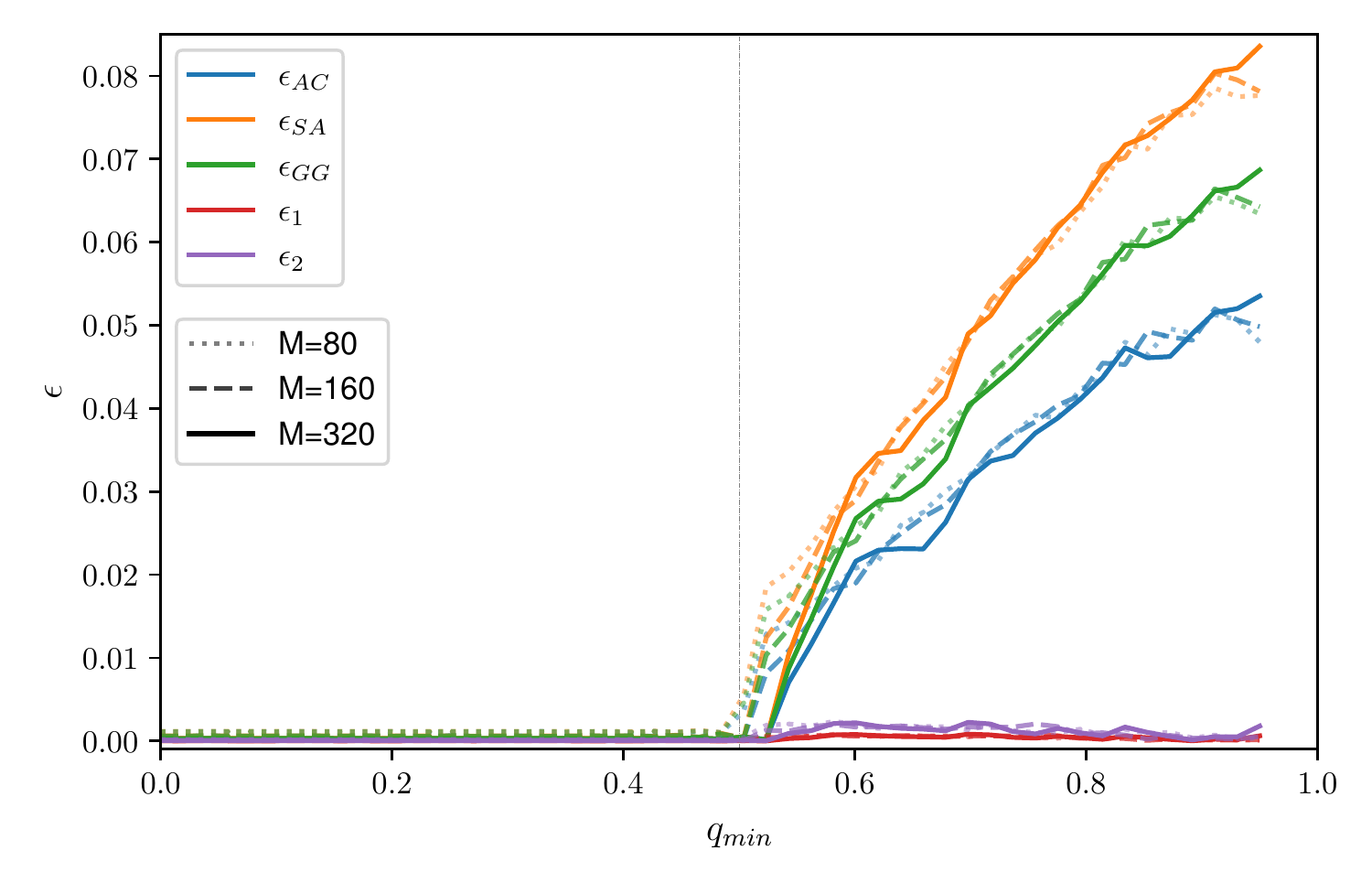}\caption{\label{Interest}Errors measured for the various ultrametric identities as a function of the mating threshold $q_{min}$ for the SFM and for different values of $M$; the process average is taken over a run of $10^2 \times M$ generations. At the numerical level, the new set of identities keep holding also for the SFM, while all the other constraints seem to be violated.}
\end{figure}

\subsubsection{Numerical inspection of ultrametric constraint's violation in the SFM}
We now study numerically the validity of all the ultrametric identities as well as of the self-averaging, by measuring the related errors $\varepsilon$. Specifically, we set $\lambda =1$ in such a way that the expected value in the HPM is $\overline {\langle q \rangle} = \frac{1}{2}$ and we vary $q_{min} \in [0,1]$. We simulate the evolution over a population of size $M$ and a time span $10^2 \times M$, and we collect data for $\varepsilon_{1},\ \varepsilon_{2},\ \varepsilon_{\textrm{GG}},\ \varepsilon_{\textrm{SA}},\ \varepsilon_{\textrm{AC}}$ that are plotted in Figure \ref{Interest} versus $q_{min}$. 
As expected, when $q_{min} <\frac{1}{2}$, the threshold does not involve significant effects with respect to the HPM and a replica-symmetric scenario is recovered with errors $\varepsilon$ close to zero. Conversely, the region $q_{min} > \frac{1}{2}$ is non-trivial and there emerge differences between the errors. As for the classical identities and for the variance, the related errors ($\varepsilon_{\textrm{SA},\textrm{GG},\textrm{AC}}$) are non-vanishing, and their values grows with $q_{min}$ without any robust trend with respect to the size $M$; as for the new identities, the related errors ($\varepsilon_{1,2}$) remain close to zero.

\subsubsection{Numerical inspection of ultrametric constraint's violation in biological and artificial human genomes}
In this Section we look for any evidence of the relations discussed before in actual genomic datasets. To this aim we tested all the ultrametric identities and the self-averaging property on the biological genome collected by {\em The 1000 Genomes Project Consortium}  \cite{Dati1} and on artificial genomes generated by two neural neworks (a Generativa Adversarial network and a Restricted Boltzmann machine) \cite{Dati2}; notably, the latter have already proved to reproduce correctly allele frequencies, linkage disquilibrium, pairwise haplotype distances and population structure. 

As in \cite{Dati2,Dati3}, we consider a population of $M=2504$ individuals  ($\sim 5 \cdot 10^3$ haplotypes) spanning $N=805$ Single Nucleotide Polymorphism (SNPs)\footnote{Single nucleotide polymorphisms are the most common type of genetic variation among people: each SNP represents a difference in a single nucleotide (e.g., a SNP may replace the nucleotide cytosine $C$ with the nucleotide thymine $T$ in a certain stretch of DNA).} from \cite{Dati1}, which reflect a high proportion of the population structure present in the whole dataset \cite{Dati2,25}. The various fluctuation relations are
evaluated by splitting the dataset of $M$ individuals into $\sqrt{M}$
groups: the population average $\langle\:\cdot\:\rangle$ is carried
out by identifying distinct replica indices with distinct individuals
within the same group. In contrast, the process average $\overline{\:\cdot\:}$
is carried out by performing an arithmetic mean over the different
evaluations of each group. Regarding the finite size scaling in $N$ it has 
been carried out by selecting a common subset of size $N$ of the genome 
variable $(-1, +1)$ for each individual.

\begin{figure}
\centering{}\includegraphics[scale=0.80]{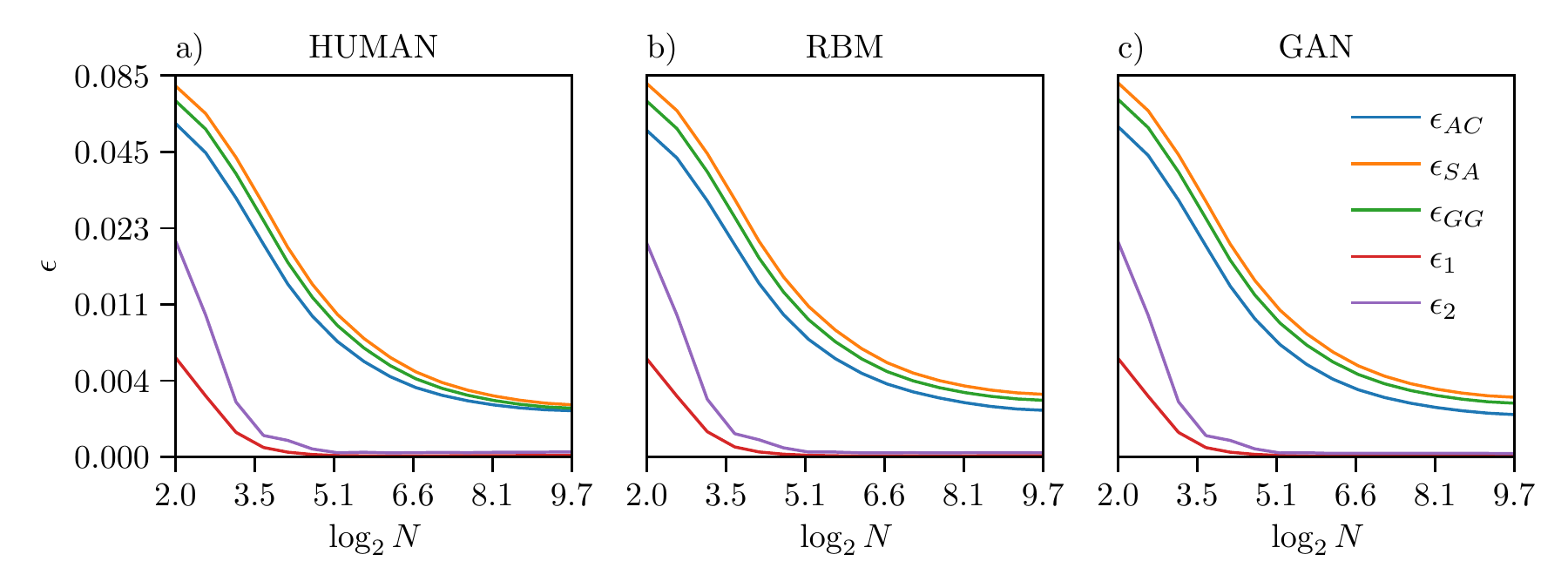}
\caption{\label{genoma} Finite-size-scaling on genome length testing the various ultrametric identities for the real human genome (HUMAN, taken from \cite{Dati1}) and two artificial genomes (taken from \cite{Dati2}) generated rispectively with a Generative Adversarial Network (GAN) and a Restricted Botlzmann machine (RBM). In these settings still our new family of identities seems to be respected, while mild violations of the others persist (despite their violation is minimal, i.e. $\varepsilon_{\textrm{AC}} \sim \varepsilon_{\textrm{GG}} \sim O(10^{-3})$).}
\end{figure}

Results are collected in Figure \ref{genoma} and show that also in these structured datasets the new set of identities is better respected w.r.t. the classical ones (despite the violation of the latter is minimal in these settings).  

%\begin{figure}[!]
%\centering{}\includegraphics[scale=0.9]{SFM.pdf}\caption{\label{Interest}The various values of the violation of the constraints $\epsilon$ (see the legend for interpretation of the colour and tickness of the curves) as a function of $q_{min}$ for different values of $M$. At the numerical level, the new set of identities keeps holding also for the SFM, while all the other constraints seem to be violated. Eventually the finite size scaling in $M$ suggests a very moderate decrease of their violation for the Ghirlanda-Guerra and Aizenman-Contucci identities, but not for the self-averaging, as expected in a broken replica scenario.}
%\end{figure} 

\section{Conclusions}

%Once made a connection among the tradeoff between replication efficiency and frequency of mutations in Darwinian evolution and the compromise between energy minimization and entropy maximization in statistical mechanics,  introduced the concept of {\em error threshold} in the mutation rate to mimic also the thermal noise (i.e. the temperature)  soon became clear that the rugged valleys of spin glass landscapes were optimal theoretical banchmarks where testing evolutionary models. To stress even more the analogy, in both these scenarios lies the presence of a double average when evaluating fluctuating quantities, i.e. over the population and over the process in the former vs thermal and quenched in the latter.   

The non-self-averaging behavior of the order parameter in spin-glass models is a peculiar,  intensively-studied feature which can be described in terms of a set of relations connecting the fluctuations of the order parameter. 
Driven by strong analogies between Natural Evolution and statistical mechanics of disordered systems, we investigated the validity of these ultrametric relations and the existence of other kinds of relations focusing on three stochastic models of evolving populations in flat landscapes. These models are the fairly standard ones in the Literature on Natural Evolution without selective pressure, that is (i) the One Parent Model (OPM) -- where reproduction is asexual and the distribution of genetic distances lacks self-averaging -- (ii) the Homogeneous Population Model (HPM) -- where reproduction is sexual and with random mating (i.e., regardless the genetic distance) and thus results in a replica symmetric picture where the genetic distance between pairs of individuals has vanishing fluctuations in the thermodynamic lmiit (hence the adjective {\em homogeneous} in the model's name) -- and (iii) the Specie Formation model (SFM) where reproduction is still sexual but with a threshold on the required similarity among mating genomes before duplication. The latter represents the most interesting case as it is the closest to biology and it spontaneously gives rise to a complex dynamics reaching a steady state with new species that are continuously and spontaneously generated and suppressed during the evolutionary process. Further, while in the first model the evolutionary tree is assumed and it is related to single descendants from common ancestors, in the latter the evolutionary tree emerges and it works at the level of species rather than single elements. 
%further the two types of ultrametricity shown by the first (OPM) and last (SFM) models are deeply different as, while the ultrametricity in the former is assumed by construction, ultrametricity in the latter is an emergent phenomenon.
\newline
Focusing on fluctuations in the genetic distances between individuals, as far as the OPM is concerned, after checking that self-averaging is absent in this model statistics, we have shown by a finite-size-scaling argument  that nor the Ghirlanda-Guerra identities neither the Aizenman-Contucci polynomials are respected. On the other hand, we were able to prove a new class of identities that are indeed respected also in our finite-size numerical checks. For HPM, as it is replica symmetric, all these constraints are equally guarantee to converge to zero in the asymptotic limit but they do not convey actual information. Then, dealing with the SFM, our identities continue to hold, being only mildly affected by finite-size effects.

As a final test we focused on human gemomes: we considered the real biological dataset taken from the {\em 1000 genome project consortium} and two synthetic datasets on artificial genomes generated by neural networks and, for all these three cases the scenario depicted by the SFM seems to be confirmed here as well: the new set of ultrametric identities is sharply respected while mild violations affect both Ghirlanda-Guerra identities as well as Aizenman-Contucci polynomials. 

As models of Natural Evolution under selective pressure, namely Darwinian Evolution, are known to display standard Ghirlanda-Guerra fluctuations (from the Franz-Peliti-Sellitto model, i.e., the $P \to \infty$ limit of the Kauffman-Levin P-spin-glass model, or the REM in a spin-glass jargon to the equal-trap model analyzed by  Leuth\"ausser and Tarazona the Hopfield model in spin glass jargon), a similar analysis to the present one should be conducted also in these settings  to better infer the role covered by Natural Selection (beyond random mutation) in shaping evolutionary taxonomies because, at present, these new findings seem to be in better agreement with Kimura Theory of Neutral Evolution: we plan to report soon on this topic.
%\newline
%As a last remark, further efforts should be also spent in evaluating if respecting these ultrametric identities can be a possible criterion for improving fast sequence alignments, on which we also plan to report soon.

\section*{Acknowledgments}
This work is supported by Ministero degli Affari Esteri e della Cooperazione Internazionale (MAECI) through the grant {\em BULBUL} (F85F21006230001), and by The Alan Turing Institute through the Theory and Methods Challenge Fortnights event "Physics-informed machine learning", which took place on 16-27 January 2023 at The Alan Turing Institute headquarters.
\newline
The authors acknowledge financial support from PNRR MUR project PE0000013-FAIR and from Sapienza University of Rome (RM120172B8066CB0).  
\newline
The authors are indebted with Aur\'elien Decelle, Silvio Franz, Luca Peliti, and Beatriz Seoane for precious discussions.
\newline

\bibliographystyle{abbrv}   %{apalike} 

\end{document}